\documentclass[10pt,conference]{IEEEtran}
\IEEEoverridecommandlockouts
\usepackage{cite}
\usepackage{amsmath,amssymb,amsfonts}
\usepackage{algorithmic}
\usepackage{graphicx}
\usepackage{textcomp}
\usepackage{xcolor}
\def\BibTeX{{\rm B\kern-.05em{\sc i\kern-.025em b}\kern-.08em
    T\kern-.1667em\lower.7ex\hbox{E}\kern-.125emX}}
\begin{document}

\title{ADROIT6G DAI-driven Open and Programmable Architecture for 6G Networks\\
}

\author{
 \IEEEauthorblockN{
Christophoros Christophorou\IEEEauthorrefmark{1}, Iacovos Ioannou\IEEEauthorrefmark{1}, Vasos Vassiliou\IEEEauthorrefmark{1}, Loizos Christofi\IEEEauthorrefmark{2}, John S Vardakas\IEEEauthorrefmark{3}, \\ Erin E Seder\IEEEauthorrefmark{4}, Carla Fabiana Chiasserini\IEEEauthorrefmark{5}, Marius Iordache\IEEEauthorrefmark{6}, Chaouki Ben Issaid\IEEEauthorrefmark{7}, \\ Ioannis Markopoulos\IEEEauthorrefmark{8}, Giulio Franzese\IEEEauthorrefmark{9}, Tanel Järvet\IEEEauthorrefmark{10}, Christos Verikoukis\IEEEauthorrefmark{11}\\
}
\\

\IEEEauthorblockA {\IEEEauthorrefmark{1} CYENS - Centre of Excellence, Cyprus }
\IEEEauthorblockA{\IEEEauthorrefmark{2} eBOS Technologies Limited, Cyprus }

\IEEEauthorblockA{\IEEEauthorrefmark{3} Iquadrat Informatica SL, Spain }

\IEEEauthorblockA{\IEEEauthorrefmark{4} Nextworks S.R.L., Italy }

\IEEEauthorblockA{\IEEEauthorrefmark{5} Consorzio Nazionale Interuniversitario per le Telecomunicazioni, Italy }

\IEEEauthorblockA{\IEEEauthorrefmark{6} Orange Romania SA, Romania }

\IEEEauthorblockA{\IEEEauthorrefmark{7} University of Oulu, Finland }

\IEEEauthorblockA{\IEEEauthorrefmark{8} NOVA Telecommunications Single Member SA, Greece }

\IEEEauthorblockA{\IEEEauthorrefmark{9} EURECOM, France }

\IEEEauthorblockA{\IEEEauthorrefmark{10} CAFA Tech Ltd, Estonia }

\IEEEauthorblockA{\IEEEauthorrefmark{11} Athena Research Centre, Greece}
  
}

\maketitle

\begin{abstract}
In the upcoming 6G era, mobile networks must deal with more challenging applications (e.g., holographic telepresence and immersive communication) and meet far more stringent application requirements stemming along the edge-cloud continuum. These new applications will create an elevated level of expectations on performance, reliability, ubiquity, trustworthiness, security, openness, and sustainability, pushing the boundaries of innovation and driving transformational change across the architecture of future mobile networks. Towards this end, ADROIT6G proposes a set of disruptive innovations with a clear vision on setting a 6G network architecture that can be tailored to the requirements of innovative applications and match the ambitious KPIs set for 6G networks. More specifically, the key transformations that ADROIT6G considers essential to 6G network evolution are: i) AI/ML-powered optimisations across the network, exploring solutions in the “Distributed Artificial Intelligence (DAI)” domain for high performance and automation; ii) Transforming to fully cloud-native network software, which can be implemented across various edge-cloud platforms, with security built integrally into the network user plan; and iii) Software driven, zero-touch operations and ultimately automation of every aspect of the network and the services it delivers.
\end{abstract}


\section{Introduction}
In next generation 6G mobile networks, the 5G application types will be redefined by morphing the classical service classes of massive Machine Type Communications (mMTC), Ultra-Reliable Low Latency Communications (URLLC), and enhanced Mobile Broadband (eMBB), to deal with more challenging applications (e.g., holographic telepresence and immersive communication) and meet far more stringent application requirements stemming along the edge-cloud continuum. These new applications will create an elevated level of expectations on performance, reliability, ubiquity, trustworthiness, security, openness, and sustainability, pushing the boundaries of innovation and driving transformational change across the architecture of future mobile networks. To meet these new expectations, a completely new end-to-end co-design of communication, control, and computing functionalities will be required, which has been largely overlooked to date. Seamless approaches must be found for transforming wireless systems into an autonomous and intelligent network fabric, which flexibly provisions and orchestrates communication-computing-control-localisation-sensing resources tailored to the requisite scenario.

This implies that the future mobile communications fabric will need to be architected differently to meet the emerging stringent requirements of new future-looking applications \cite{b51,b52,b53} that cannot be served by existing 5G mobile networks. The network evolution may happen in many ways. In ADROIT6G, we recognise the following two as the most prevalent:
\begin{itemize}
\item\textbf{i) Network of networks:} This concept entails numerous layers within and across networks, including macro, ultra-dense small cell, and 3GPP-compatible Non-Terrestrial Networks (NTN), working seamlessly and cohesively to provide a new level of ubiquity, performance, and reliability that emerging applications will require; and 
\item\textbf{ii) Network-as-a-Service:} Networks will be consumed dynamically and as-a-service, similar to cloud services of today; coordination between the cloud and the network is essential to successfully execute on enterprise transformations.
\end{itemize}

In ADROIT6G we are aiming at laying the foundations for long-term research for low TRL technology advancements starting with the definition of a disruptive sixth generation (6G) wireless system architecture that: \textbf{i)} has a design inherently tailored to the performance requirements of new future-looking applications and their accompanying technological trends; \textbf{ii)} will be a self-contained, secure ecosystem of fully distributed AI-powered dynamic intelligence at all network layers; \textbf{iii)} will integrate the 6G terrestrial, and NTN communications into a robust network, which would be more reliable, faster, provide higher availability and end-to-end security, and can support a massive number of devices with ultra-low latency requirements; and \textbf{iv)} can truly integrate far-reaching future-looking applications ranging from autonomous systems to extended reality. 

While ADROIT6G primarily focuses on performance, reliability, and scalability, network and end-to-end devices and services must also prioritise security and privacy. New security principles, such as zero trust, will be incorporated to ensure that security and privacy considerations do not compromise these characteristics. To address these aspects, collaboration with other security-focused initiatives will be pursued. 

Although we highly recognize the importance of quantitative results to demonstrate the efficacy of our proposed architecture, our current research efforts within the ADROIT6G are still in the initial phase and centred on theoretical development and foundation groundwork, instrumental in setting the stage for the future development of the ADROIT6G project; our emphasis has been on conceptualizing the proposed 6G architecture and defining the foundational innovative concepts and technologies driving the ADROIT6G system. Thus, while this paper extensively details our innovative concepts and envisioned transformations, it's crucial to highlight that empirical tests and quantitative validations have not been conducted yet. However, these form an integral part of our upcoming phases, where practical implementations and extensive evaluations will be conducted and included in subsequent works. Specifically, for validating the ADROIT6G concept and technologies, a blend of theoretical considerations, simulations, and emulations will be utilized alongside upgraded 5G testbeds. Components that are not or are not easily available (e.g., satellite-based edge clouds, traffic generators) will be emulated or simulated. The ADROIT6G integrated system will be tested and validated in a lab environment (TRL4), ensuring its effectiveness and reliability.

Following this introductory Chapter, the rest of the paper is structured as follows: Chapter \ref{Vision} describes the vision and key transformations that ADROIT6G considers essential to 6G network evolution. Chapter \ref{Architecture} describes the overall ADROIT6G conceptual architecture. The innovative concepts and technologies driving
the ADROIT6G system are described in Chapter \ref{InnovativeConcepts}. Finally, Chapter \ref{Conclusions} provides the Conclusions.

\begin{figure*}
	\begin{center}
	\resizebox{\linewidth}{!}{%
	\includegraphics[width=\textwidth]{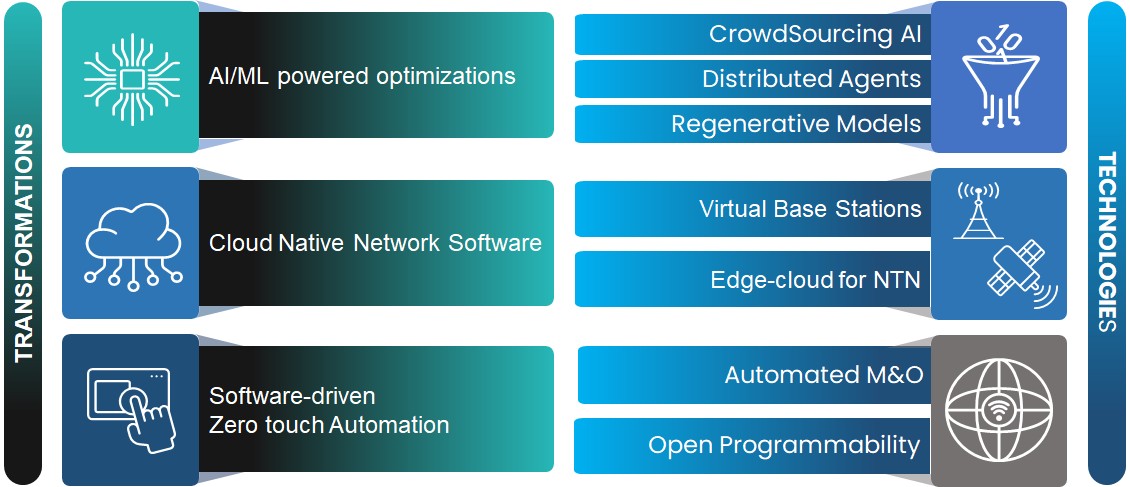}
	}
	\caption{ADROIT6G Vision and Key Transformations towards 6G Network Evolution}
\label{visualkey}
	\end{center}
\end{figure*}

\section{Vision and Key Transformations}
\label{Vision}

ADROIT6G proposes a set of disruptive innovations with a clear vision on setting a 6G network architecture that can be tailored to the requirements of innovative applications and meet the ambitious KPIs set for 6G networks. The key transformations (see Fig. \ref{visualkey}) that ADROIT6G considers essential towards this 6G network evolution, are:  

\begin{enumerate}
  \item \textbf{AI/ML-powered optimisations across the entire network, for high performance and automation.} In ADROIT6G, we are moving from the “Standard” Centralised AI model towards exploring solutions in the “Distributed AI” domain, both with Federated Learning and Decentralised Learning with the use of: \textbf{i) CrowdSourcing AI framework:} This novel concept of Crowdsourcing AI is introduced to serve the different domains and applications with the “best” AI models for their decision-making processes. It aims to improve energy efficiency, minimise the environmental carbon footprint and enable secure and efficient AI/ML operations in distributed systems; \textbf{ii) Distributed Agents in the form of Believe-Desire-Intention (BDI) agents:} This method implements a truly decentralised learning form and can operate anywhere in the edge-cloud continuum; and \textbf{iii) Distributed representation learning of application data:} The design of generative models from available multimodal data allows for the saving of resources.
         
  \item \textbf{Transforming to a fully cloud-native network software, which can be implemented across various edge-cloud platforms, with security built integrally into the network user plane. }In ADROIT6G, we are evolving cellular infrastructure, and we are making its operation cloud-native, with the use of: \textbf{i) User Equipment as Virtual Base Stations (UE-VBSs)}, enabling the deployment of networking components at the far edge; and ii) \textbf{Edge Cloud Deployments in relation to NTN,} for true cellular and satellite communications integration.
          
  \item \textbf{Software-driven, zero-touch operations, and ultimately automation of every aspect of the network and its services.} Part of the challenge is that it is not easy to foresee all the areas where new solutions will be needed, so it is essential to have a model that allows for on-the-fly collaboration and adaptation. In ADROIT6G, we are evolving the Management \& Orchestration (M\&O) into a fully automated M\&O solution, via the distribution of all the M\&O functions and the use of the Distributed AI (DAI) framework, for an optimised closed-loop control.
\end{enumerate}

Specifically, the AI/ML-driven operation enables scalability and massive digitisation, including “tailor-made” implementation of various tenants’ requirements. The cloud-native environment will enable seamless operations and service execution across various heterogeneous infrastructures, domains, services, business, and application domains, with reliability and security. The software-driven, low-touch/zero-touch operation supports a consistent/reliable programmable environment. 

An initial design of the ADROIT6G conceptual architecture, as well as the main innovative concepts and technologies driving the ADROIT6G system, are provided in Chapter \ref{Architecture} and Chapter \ref{InnovativeConcepts}, respectively. 

\section{ADROIT6G Conceptual Architecture}
\label{Architecture}

\begin{figure*}
	\begin{center}
	\resizebox{\linewidth}{!}{%
	\includegraphics[width=\textwidth]{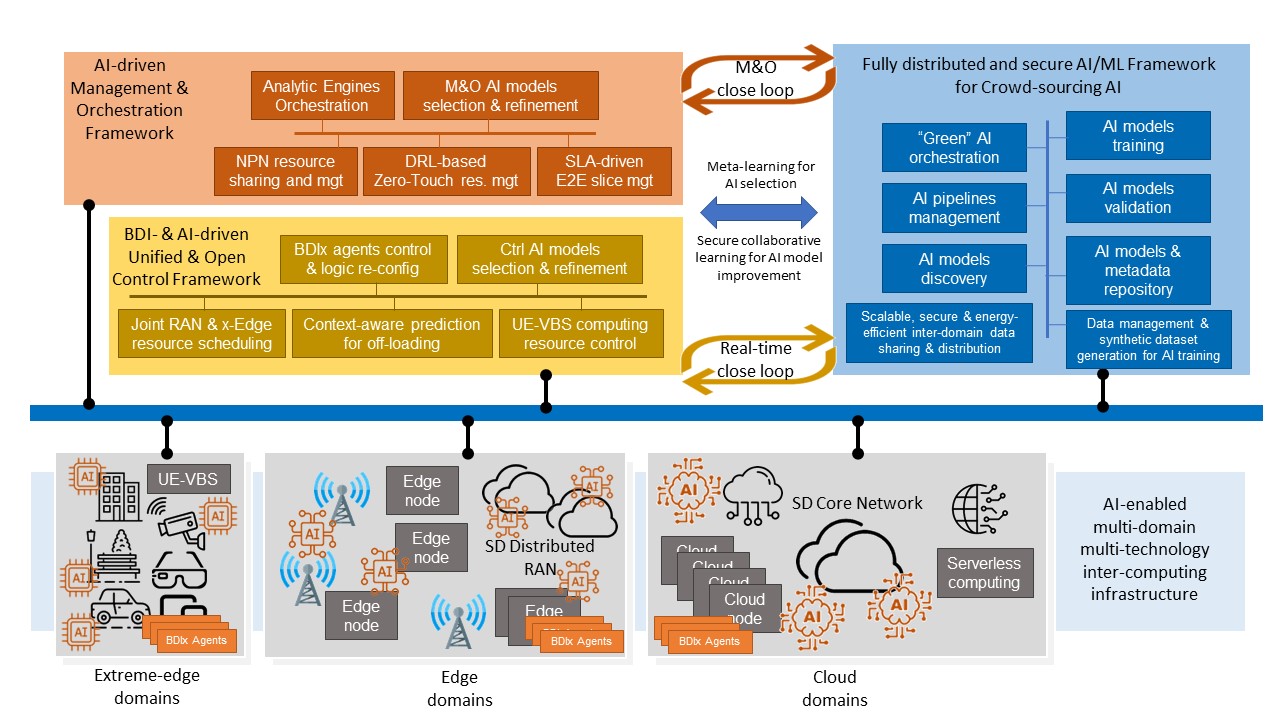}
	}
	\caption{ADROIT6G Conceptual Architecture}
\label{ADROIT6Garchitecture}
	\end{center}
\end{figure*}

The ADROIT6G architecture evolves the existing architecture of 5G networks, by adopting a fully distributed and dynamic paradigm, with functional elements automatically deployed on demand as virtual functions in cloud-native environments across far-edge, edge, and cloud domains operated by different stakeholders. As illustrated in Fig. \ref{ADROIT6Garchitecture}, the ADROIT6G architecture consists of three cooperative inter-domain frameworks that use over a programmable inter-computing and inter-network infrastructure. The distributed computing nodes at the far edge, edge, and cloud domains, each of them with their own characteristics and capabilities, are used to deploy virtual functions of software-defined disaggregated RAN and core network, virtual applications, as well as AI agents, which are orchestrated dynamically as part of the overall network control and management strategies.

The three cooperative blocks/frameworks of the ADROIT6G architecture, dedicated to M\&O, Control, and AI-driven Management, respectively, provide the logic and the procedures for zero-touch network management and control. More specifically:

\begin{itemize}
  \item The \textbf{AI-driven Management and Orchestration framework}, implements the logic for the end-to-end, automated management of the inter-computing and inter-domain network infrastructure. It includes features for the composition of multi-tenant network slices across edge, transport, and core domains. This allows for coordinating the sharing of the slice subnets, the lifecycle of RAN, core network and MEC virtual functions, and the resource allocation. These are in compliance with the established Service Level Agreements (SLAs), the required levels of isolation and the dynamic QoS demands of the running services. The slice management is extended for scenarios that integrate public and non-public networks. In this context, the ADROIT6G system will implement service-driven strategies to share or allocate dedicated user plane functions at the MEC and jointly manage the resources at the far edge.
  \item The \textbf{BDIx \& AI-driven Unified and Open Control framework}, is responsible for the real-time and near-real-time closed loops for network automation. It coordinates the dynamic and inter-domain resource allocation in the short term based on the current network condition and context-aware predictions. Examples of actions handled by the control framework are the mitigation of resource congestion, the joint scheduling of RAN and x-edge resources to decide the association of mobile devices to radio access points and edge hosts, the energy-efficient computation off-loading driven by distributed learning mechanisms, and the joint and proactive allocation of VMs and contents in the edge to guarantee service delay constraints. The control framework is also responsible for the provisioning and control of the BDIx agents in the various domains, the dynamic re-configuration of their logic, and the control of the User Equipment as Virtual Base Station (UE-VBS) computing concept resources, exploiting the capabilities of the devices at the far edge.
  \item The\textbf{ Fully distributed and secure AI/ML framework for CrowdSourcing AI}, provides the AI mechanisms to assist the M\&O and the control frameworks in their decisions. The interaction is bidirectional. The AI/ML framework offers the mechanisms for energy-efficient provisioning of AI agents, initial training of AI models, and their discovery and selection based on the concept of meta-learning. On the other hand, the M\&O and Control frameworks support the continuous validation and refinement of the AI models towards secure and collaborative learning for AI models’ improvement. The AI/ML framework also implements the data management and distribution mechanisms, with: i) procedures to share information across multiple domains in a secure and scalable manner; and ii) mechanisms to generate synthetic datasets for AI training in scenarios with scarce, incomplete, or low-quality data.
\end{itemize}

These are based on advanced ML techniques and Crowdsourcing AI, and adopt a cooperative and fully distributed model. Monitoring, analysis, decision, and execution actions are implemented by distributed functions that operate at different layers (physical infrastructures, network slices, and services) and in the scope of different domains. Multiple real-time and medium/long-term close loops are executed in parallel and coordinated together to guarantee the overall stability and consistency of the global allocation strategy, the real-time reaction to unexpected conditions, and the continuous resource optimisation in the end-to-end network. 

A common, secure, and scalable inter-domain communication bus enables their interaction, regulating the exposure of the capabilities offered and consumed by the cooperating domains, enabling a dynamic publish/subscribe of APIs and resources, and integrating access control and authorisation mechanisms. Following this paradigm, the architectural components are loosely coupled with the others. Still, they can collaborate using the concept of exposed services, which can be dynamically advertised, discovered, and consumed through open interfaces, e.g., in the form of REST APIs. 

Moreover, ADROIT6G architecture evolves the well-known concept of ETSI NFV MANO framework, with an orchestrator that coordinates the provisioning of network services, instantiating virtual functions over a potentially distributed virtual infrastructure and enabling their interaction through a set of well-defined virtual links interconnecting their connection points. ADROIT6G proposes a distributed orchestration and control logic that manages dynamic meshes of container-based service and application components. Moreover, on its own, such orchestration logic is structured in a mesh of cooperating functions where, in turn, some of them can be dynamically instantiated, orchestrated, and re-configured, e.g., adopting concepts of AI-as a Service and Monitoring as a Service.

\section{ADROIT6G Innovative Concepts and Technologies}
\label{InnovativeConcepts}
ADROIT6G aims to realise a disruptive AI-powered, cloud-native, sustainable network ecosystem consisting of a novel control layer that enables domain-specific AI/ML tasks for both network management and user-oriented services. ADROIT6G also proposes revolutionary data management, generation, and transfer techniques by optimally exploiting context and semantic information. 

The  Innovative Areas (IAs) driving the ADROIT6G system can be categorized as follows:
\begin{itemize}
\item \textbf{AI/ML-motivated Innovation:} The contributions of ADROIT6G in this category come with the Distributed AI framework based on BDIx Agents (IA 2), the semantic communication approach as an alternative to Shannon theory (IA 5), the CrowdSourcing AI initiative (IA 6), and the employment of multimodal representation learning (IA 7).
\item \textbf{Computing Concept Innovation:} The contributions of ADROIT6G in this category come with the pioneering User Equipment as Virtual Base Station (UE-VBS) Computing concept (IA 1), harnessing unutilized resources in end devices to redefine cellular, NTN, and Edge Computing infrastructures.
\item \textbf{Management and Orchestration Innovation:} ADROIT6G innovations in this category come with the Zero-Touch Management design (IA 3), paired with advanced strategies for network automation, self-optimisation, and orchestration across multiple domains (IA 4).
\item \textbf{Communication Concept Innovation:} Departing from traditional norms, ADROIT6G showcases a fresh take on communication with its novel semantic communication approach (IA 5).
\end{itemize}

Below, we describe in more detail the Innovative Areas (IAs) driving this endeavour.

\subsection{IA 1: UE-VBS computing concept}
One of the key innovation areas into the ADROIT6G architecture is the User Equipment as Virtual Base Station (UE-VBS) computing concept \cite{b8,b9,b10,b11}. This concept resolves around exploiting the unutilized resources of end user Mobile Devices at the far-edge, providing a virtual distributed pool of networking, computing and storage resources, for augmenting and redefining in real-time, the cellular, Non Terrestrial Network (NTN) and edge computing infrastructure, wherever and whenever needed. It represents a major advancement in the evolution of decentralized cellular networking, that moves beyond existing paradigms by delivering an integrated solution that doesn't just involve mobile devices in the network but actively exploits their multifunctional capabilities, to tackle the extreme performance requirements of 6G use cases. Whether these are acting as virtual base stations, relay stations, or computational/storage nodes, UE-VBS Computing powered by the DAI Framework with BDIx agents,  leverages mobile devices' inherent ubiquity \cite{b5, b12} and capabilities to provide a 6G infrastructure with self-organization, self-optimisation and self-healing capabilities \cite{b101}, able to autonomously adapt cellular, edge and far-edge computing resources to the increasing complexity and volume of data, facilitating inter-computing and inter-networking across various edge, access, transport, and core components \cite{b10}.
In addition, this novel concept can address the dynamic nature and performance limitations of NTN by permitting application-specific enhancements, such as data pre-processing or cache near the UE or in peripheral cloud instances \cite{b11,b12,b13}.

\subsection{IA 2: A novel DAI framework based on BDIx Agents}
ADROIT6G introduces a novel DAI framework as part of its architecture to meet the technical requirements of 6G networks. The DAI framework concentrates on supporting applications in the class areas of mMTC, URLLC, and eMBB \cite{b101}. This framework has effectively facilitated mMTC and eMBB through efficient network management and control \cite{b102}. Incorporating a decentralised AI framework into the DAI architecture can facilitate rapid decision-making, provide self-healing mechanisms, and foster collaboration as a self-organizing network \cite{b17,b18}. The framework aims to achieve decentralisation, distributed control, and autonomy, bringing intelligence to the edge of every network terminal. It enables real-time network decisions through unsupervised learning and inter-user and inter-operator knowledge sharing. The DAI framework also incorporates self-organisation strategies, including self-learning, self-configuration, self-healing, and self-optimisation at the terminal level using AI, Deep Learning, and Machine Learning techniques. The architecture of the DAI framework is designed to be modular and allows the integration of multiple intelligent approaches. Each UE hosts a distributed, autonomous, dynamic, and flexible BDI extended (BDIx) agent with Reinforcement Learning. The framework enables the substitution or addition of beliefs and desires as modules, targeting specific tasks or requirements in 6G, such as achieving high data rates. Communication and collaborative decision-making among the BDIx agents are facilitated through message exchanges, including proposals, notifications, and information sharing.

\subsection{IA 3: Zero-Touch Management}
ADROIT6G aims at designing and implementing technology enablers for 6G network management with AI-based zero-touch automation capabilities, targeting a fully distributed and dynamic approach, with functional elements automatically deployed on-demand as cloud-native virtual functions across multi-stakeholder domains. This will be applied to the ADROIT6G architecture by leveraging on existing solutions, such for example the Zero-Touch Decision engine developed in MonB5G project \cite{b25}. To this end, a central concept is identified in the design and implementation of distributed AI agents (e.g. for Reinforcement Learning - RL) automatically deployed as cloud-native functions at all domains of the network (Core, edge, RAN and extreme edge). Moreover following the ETSI ZSM closed-loop approach and principles, a logically centralized governance engine is expected to be responsible for the coordination of the various distributed AI agents. The main aim of the governance engine is to monitor the performance and behaviour of the AI agents and will and arbitrate the decision process, considering cross-domain trade-offs or even modify Deep Reinforcement Learning (DRL) objective functions to adapt to different operating conditions and performance objectives.

\subsection{IA 4: Network automation and self-optimisation via closed-loop orchestration across multiple domains}
The ADROIT6G strategy seeks to improve the network's performance and robustness by leveraging the complete distribution of Management and Orchestration (M\&O) functions across multiple domains. This strategy employs AI engines and BDIx agents to facilitate cooperation and communication between these dispersed elements. Horizontal (per-domain) and vertical (at infrastructure, network, and service levels) scopes are considered for re\-optimisation decisions. The agents use agent communication language, such as FIPA ACL \cite{b30}, for intercommunication utilising messages and BDIx control logic. This assures the overarching consistency and coordinated execution of the end-to-end strategy. The orchestrator can update the agents' BDIx control logic and active intentions using REST services and Federated Learning. This permits continuous refinement and adaptation of the control logic in response to changing network conditions. The approach reduces potential instability by coordinating short-term control with medium- and long-term re\-optimisation actions. It ensures a high level of robustness in the face of the extreme volatility of cutting-edge resources.

\subsection{IA 5: A novel approach to semantic communication}
ADROIT6G introduces a novel approach to semantic communication that departs from the conventional Shannon-theoretic approach. Rather than focusing on precise bit reconstruction, emphasis is placed on recovering the underlying meaning of the transmitted sequence or completing the intended task. This technique uses generative models to regenerate transmitted data during data loss, accurately.  Encoders and decoders for the transmitter and receiver are designed to utilise generative models, dynamically resolving whether to transmit raw data or model parameters. The objective is to assure semantic equivalence between transmitted and received data, rather than bit-level identity. This viewpoint allows for the transmission of fewer bits while preserving semantic communication. Instead of concentrating on bit-by-bit recovery, encoding strategies can be tailored to attain specific objectives in goal-oriented situations. ADROIT6G expands on \cite{b33} by employing the information bottleneck principle to enhance compression levels based on the accuracy of objective fulfilment. The project addresses issues such as designing accurate yet compressible generative models, developing mechanisms to determine the benefit of transmitting the model or data, and establishing feedback mechanisms for evaluating semantic coherence between regenerated and transmitted data.

\subsection{IA 6: CrowdSourcing AI}
ADROIT6G intends to develop CrowdSourcing AI, a collaborative concept referred to as ML as a Service (MLaaS), in order to resolve the challenge of sustainable AI in distributed network systems. This concept enables the sharing and utilisation of ML models trained within network domains using distributed approaches such as federated learning. The CrowdSourcing AI framework enables, ensemble and transfer learning by employing meta-learning techniques to select the most appropriate ML model for specific tasks across domains. Meta-learning solutions will be implemented to optimise the trade-off between energy consumption and decision quality. By transmitting knowledge from complex models to simplified ones without the need for extensive data sharing, techniques such as knowledge distillation will be used to reduce energy consumption during inference. In addition, model quantisation and pruning will be investigated to improve on-device ML by considering inference latency, reliability, and energy efficiency. ADROIT6G will establish a distributed control layer that facilitates cooperative and energy-efficient AI in order to implement the CrowdSourcing AI concept. This control layer will facilitate efficient data and ML model sharing among system components, interacting with any mobile network platform domain or entity. The innovations in this area are anticipated to result in an enhancement of one magnitude in ML energy consumption and a reduction of fifty percent in inference energy consumption.

\subsection{IA 7: Multimodal Representation Learning for next generation 6G wireless networks}
ADROIT6G aims to simplify data collection in AI models and data-driven ecosystems by leveraging multimodal representation learning. The project will develop models to merge diverse data sources into a latent space capable of data reconstruction and downstream ML tasks like classification. These models will also be able to generate missing modalities or data, compensating for potential malfunctions and reducing bandwidth consumption. The project will focus on distributed techniques, enabling the learning of these models across far-edge, edge, and cloud nodes. The joint multimodal representation will be constructed at the edge-cloud interface. Significant advancements in multimodal generative modelling \cite{multim} and integrating these models with the ADROIT6G orchestration framework are required to achieve these objectives. This integration will enable the network infrastructure to benefit from cross-generation capabilities, reducing extreme network utilisation and improving performance.

\section{Conclusions}
\label{Conclusions}
The ADROIT6G project is a forward-thinking initiative with aim to create a sustainable, AI-powered, cloud-native 6G Architecture. Seven innovative areas form the backbone of this project, each introducing a groundbreaking concept, technology, or framework. The introduction of the UE-VBS computing concept, exploiting the resources of end-user devices, is a major stride towards enhancing the capabilities of cellular, NTN, and Edge Computing infrastructures. Simultaneously, incorporating in 6G networks a novel DAI framework based on BDIx agents, is envisaged to substantially impact the network's operational efficiency. The ADROIT6G project is also a pioneer in proposing Zero-Touch management that enables AI-driven dynamic slice reconfiguration for self-driven 6G infrastructures, ensuring smooth network operations. To further improve the network's performance and robustness, the ADROIT6G project includes network automation and self-optimization via closed-loop orchestration in multi-layer and inter-computing scenarios with multiple cooperating stakeholders. Regarding resource usage, the project presents an innovative approach to semantic communication, aiming to support new, highly demanding network services and user applications with fewer resources. The project also takes significant strides to minimise the AI/ML carbon footprint and enable efficient AI/ML in distributed systems. The CrowdSourcing AI framework is a significant innovation in this regard. Lastly, the project looks to revolutionise data usage and generation through learning representations of data collected at the edge and the far edge of the network. It adopts multimodal representation learning to streamline data collection in AI models and data-driven ecosystems.

Note that during the lifecycle of ADROIT6G, new innovative concepts and technologies may succeed stemming the dynamic and fast-paced nature of technological innovation in the field of wireless communications. As 6G continues to evolve, we can expect new research findings, technological breakthroughs, or industry trends to emerge that may influence the ADROIT6G course of development and, thus, the identified innovative concepts. Therefore, the ADROIT6G conceptual architecture as well as the list of innovative concepts and technologies described, is not fixed and could be expanded or refined over the project's lifetime, ensuring the project's outcomes stay at the cutting edge of the 6G field. These updates will be included in the next versions of this article.

\section*{Acknowledgment}
This work has received funding from the European Union under grant agreement No 101095363. However, the views and opinions expressed are those of the author(s) only and do not necessarily reflect those of the European Union. Neither the European Union nor the granting authority can be held responsible for them.

We would also like to extend our sincere appreciation to our respected partners, whose significant contributions, equal in importance to those of the authors, have been instrumental in developing this paper. The team consisting of Elli Symeou (eBOS), Marios Sophocleous (eBOS), Mehdi Bennis (UOULU),   Giacomo Bernini (Nextworks), Konstantinos Oikonomou (NOVA), Pietro Michiardi  (EURE), Antonella Molinaro (CNIT), Sergio Barbarossa (CNIT), and Stefania Sardellitti (CNIT) has together contributed their experience and commitment towards the advancement of the "ADROIT6G DAI-driven Open and Programmable Architecture for 6G Networks". The collective endeavors and perspectives of the individuals involved have greatly enhanced the substance and conceptualization of this proposal. We are grateful for their steadfast support and dedication to propelling the 6G network advancement boundaries.

\vspace{12pt}


\begin{thebibliography}{00}

\bibitem{b51}  Salameh, A.I.; El Tarhuni, M., “From 5G to 6G—Challenges, Technologies, and Applications”, Future Internet 2022.
\bibitem{b52}	Shimaa A. Abdel Hakeem, Hanan H. Hussein, HyungWon Kim, "Vision and research directions of 6G technologies and applications, Journal of King Saud University - Computer and Information Sciences, 2022.
\bibitem{b53} Asghar, M.Z.; Memon, S.A.; Ham¨ al¨ ainen, J. Evolution of Wireless Communication to 6G: Potential Applications and Research Directions. Sustainability 2022.
\bibitem{b5} D. Miller, "The Sharing Economy and How it Is Changing Industries," The Balance Small Business, 2019. [Online]. Available: https://www.thebalancesmb.com/the-sharing-economy-4178831. (Accessed: Jul. 3, 2023).
\bibitem{b8}  C. Christophorou et al., "CelEc Framework for Reconfigurable Small Cells as Part of 5G Ultra-Dense Networks," in Proc. IEEE ICC 2017 Mobile and Wireless Networking (ICC'17 MWN), Paris, France, May 2017, pp. 935-941.
\bibitem{b9}  P. Swain et al., "Selection of UE-based Virtual Small Cell Base Stations using Affinity Propagation Clustering," in Proc. 14th International Wireless Communications and Mobile Computing Conference (IWCMC 2018), Limassol, Cyprus, June 2018.
\bibitem{b10}  K. Venkateswararao et al., "Dynamic selection of Virtual Small Base Station in 5G Ultra-Dense Network using Initializing Matching Connection Algorithm," in Proc. 2019 IEEE International Conference on Advanced Networks and Telecommunications Systems, 2019.
\bibitem{b11}  K. Venkateswararao et al., "Using UE-VBS for Dynamic Virtual Small Cells Deployment and Backhauling in 5G Ultra-Dense Networks," Elsevier Computer Networks Journal (COMNET), vol. 189, Apr. 2021. doi: 10.1016/j.comnet.2021.107926.
\bibitem{b12}   Small Cell Forum, "Solving the HetNet Puzzle: Small cell siting and deployment challenges in hyperdense networks," Release 9, 2017.
\bibitem{b13} Groupe Spéciale Mobile Association (GSMA), "The Mobile Economy 2020," London, United Kingdom, 2020.
\bibitem{b101}  I. Ioannou, V. Vassiliou, C. Christophorou, and A. Pitsillides, "Distributed Artificial Intelligence Solution for D2D Communication in 5G Networks," IEEE Systems Journal, vol. 14, no. 3, pp. 4232-4241, 2020. doi: 10.1109/JSYST.2020.2979044.
\bibitem{b102}  I. Ioannou, C. Christophorou, V. Vassiliou, and A. Pitsillides, "Performance Evaluation of Transmission Mode Selection in D2D communication," in Proc. 11th IFIP International Conference on New Technologies, Mobility and Security (NTMS), 2021.
\bibitem{b17}  "5G D2D Transmission Mode Selection Performance \& Cluster Limits Evaluation of Distributed AI and ML Techniques," August 2021, doi: 10.1109/COMNETSAT53002.2021.9530792.
\bibitem{b18}  I. Ioannou, C. Christophorou, V. Vassiliou, and A. Pitsillides, "Distributed Artificial Intelligence Solution for D2D Communication in 5G Networks," April 2020.
\bibitem{b25}  A. Dalgkitsis et al., "SCHEMA: Service Chain Elastic Management with Distributed Reinforcement Learning," in IEEE Global Communications Conference (GLOBECOM), 2021.
\bibitem{b26}  S. Chen et al., "Stabilization Approaches for Reinforcement Learning-Based End-to-End Autonomous Driving," in IEEE Transactions on Vehicular Technology, vol. 69, no. 5, pp. 4740-4750, May 2020.
\bibitem{b30}  http://www.fipa.org/repository/aclspecs.html

\bibitem{b33}  F. Pezone, S. Barbarossa, P. Di Lorenzo, “Goal-oriented communication for edge learning based on the information bottleneck principle”, ICASSP 2022.

\bibitem{multim} M. Bounoua, G. Franzese, P. Michiardi, "Multi-modal Latent Diffusion", arXiv preprint arXiv:2306.04445 (2023).

\end{thebibliography}
\end{document}